\documentclass{PoS}

\title{Euclidean partons?}

\ShortTitle{Euclidean partons?}

\author{\speaker{G.C. Rossi}
\\
Universit\`a di Roma Tor Vergata \& INFN - Sezione di Roma Tor Vergata\\
Via della Ricerca Scientifica, 1 - 00133 Roma, ITALY\\
and\\
Centro Fermi - Museo Storico della Fisica e Centro Studi e Ricerche 
 ``E.\ Fermi''\\
 Piazza del Viminale, 1 - 00184 Roma, ITALY\\ 
        E-mail: \email{rossig@roma2.infn.it}}
        \author{M. Testa\\
Sapienza Universit\`a di Roma \& INFN - Sezione di Roma \\
Piazzale A. Moro, 2 - 00185 Roma, ITALY\\
        E-mail: \email{massimo.testa@roma1.infn.it}}

\abstract{In this talk we reexamine the possibility of evaluating parton distribution functions from lattice simulations. We show that, while in principle individual moments can be extracted from lattice data, in all cases the process of renormalization, hindered by lattice momenta limitation, represents an obstruction to a direct calculation of the full parton distribution function from QCD simulations. We discuss the case of the Ji quasi-parton distribution functions, the possibility of using the reduced Ioffe-time distributions and the more recent proposal of directly subtracting power divergent mixings in perturbation theory.}

\FullConference{XIII Quark Confinement and the Hadron Spectrum - Confinement2018\\
		31 July - 6 August 2018\\
		Maynooth University, Ireland}

\begin{document}

\section{Introduction}
\label{sec:INTRO}

In the seminal paper of ref.~\cite{Ji:2013dva} it was suggested that parton distribution functions (PDFs) could be non-perturbatively computed via numerical lattice QCD simulations making use of the formula~\footnote{To reduce the discussion to its essentials and avoid irrelevant kinematical complications we drop all flavour and Lorentz indices on the hadronic currents. We shall then consider a hypothetical theory of ``scalar quarks'' in which an appropriately renormalized scalar current, $j(x)=:\phi(x)\phi(x):$, carrying momentum $q$ ($q^2 < 0$), hits a scalar ``proton''.}
\begin{eqnarray}
F (\omega) = \lim_{P \rightarrow \infty} \frac {P}{2 \pi} \int_{-\infty}^{+\infty} d z\, e^{iz \omega P} \langle P|\phi (0) \phi(\xi)|P\rangle\Big{|}_{\xi=(0,0,0,z)} \, , \label{proposal}
\end{eqnarray}
where $|P\rangle$ denotes the state of a proton with momentum $P$ along the (negative) $z$-axis and $\xi$ is the space-time event $(0,0,0,z)$. Eq.~(\ref{proposal}) expresses $F(\omega)$ in terms of the matrix element of an  ``equal-time'' (ET) operator which thus takes the same value in Minkowski and in Euclidean metric. Consequently its computation can be in principle performed in lattice  simulations. A lot of theoretical and numerical work was then invested in trying to give a firm and workable basis to the above formal equation (see the many references in the community white book~\cite{Lin:2017snn}). 

Although in the formal $P\to \infty$ limit the structure function~(\ref{proposal}) obtained from the ET bilocal 
\begin{equation}
{\cal M}^{ET}(P\cdot \xi,\xi^2=-z^2)=\langle P| \phi (0) \phi(\xi)|P\rangle \Big{|}_{\xi=(0,0,0,z)} 
\label{ETB}
\end{equation}
equals the one obtained from the matrix element of the light-cone (LC) bilocal 
\begin{equation}
{\cal M}^{LC}(P\cdot \xi,\xi^2=0)=\langle P|\phi (0) \phi(\xi)|P\rangle \Big{|}_{\xi=(z,0,0,z)} \, ,
\label{ELC}
\end{equation}
it has been shown in refs.~\cite{Rossi:2017muf,Rossi:2018zkn} that the ET    formula~(\ref{proposal}) cannot provide the correct expression of the PDF because its moments are power divergent~\footnote{We stress that we are not talking here about the power divergencies associated with the exponentiating linear mass-like divergency associated with the presence of the Wilson line between $\bar q$ and $q$ in the QCD bilocal.}. Such power divergencies come from the mixing of deep inelastic scattering (DIS) operators with lower dimensional ones ({\it trace} operators) and cannot be cured (eliminated) by any multiplicative renormalization that may be employed to make finite the logarithmically divergent quantity ${\cal M}^{ET}(-P\, z,-z^2)$. Since the PDF moments are instead physically measurable (and measured) observables~\cite{Lai:1996mg,Bluemlein:2002be,Christy:2012tk}, one must conclude that it is not possible to directly extract PDFs from lattice simulations based on the formula~(\ref{proposal}). 

These considerations highlight an obstruction to a non-perturbative calculation of the PDF from Euclidean data. However, they do not prevent~\cite{Dawson:1997ic} extracting single PDF moments from short distance OPE expansions like those considered in refs.~\cite{Karpie:2018zaz} or~\cite{Ma:2017pxb}.

The outline of the talk is as follows. We start in sect.~\ref{sec:PBO} by explicitly proving that, neglecting renormalization effects, in the formal $P\to \infty$ limit the ET Ji formula~(\ref{proposal}) yields the Minkowski LC structure function. In sect.~\ref{sec:RENMAT} we show that the PDF moments derived from the renormalized relation~(\ref{proposal}) are UV power divergent, so the latter cannot be used to evaluate the PDF from lattice data. In sect.~\ref{sec:RIT} we discuss the difficulties with the recent interesting proposal of subtracting the UV power divergent mixings in perturbation theory (PT)~\cite{Radyushkin:2018nbf}, We give in sect.~\ref{sec:CONC} a few concluding remarks. In Appendix we discuss a toy-model in which we show that UV divergent mixings with trace operators may not show up as (power) divergencies in the PDF but have the effect of deforming its expression.

\section{Partons and bilocal operators}
\label{sec:PBO} 

In this section we want to show that, neglecting renormalization effects, the Ji proposal for the structure function gives back the LC structure function. The situation changes radically if renormalization issues are properly accounted for. In sect.~\ref{sec:RENMAT} we describe the modifications occurring when the theory is not canonical, separately analyzing the Minkowski and Euclidean case. We conclude that the moments of the structure function associated with the hadronic matrix elements of the Ji bilocal operator are plagued with UV power divergences. This represents an obstruction to a direct lattice calculation of the PDF using the formula~(\ref{proposal}).

\subsection{Deep inelastic Scattering in the parton model}
\label{sec:DISSF}

For completeness we start by recalling how $\langle P| \phi (0) \phi(\xi)|P\rangle$ is related to the DIS cross section and how DIS operators are defined in the canonical parton model.

\subsubsection{DIS cross section and parton structure function}
\label{sec:CANC}

In the canonical parton model the bilocal is a regular function of $\xi^2$ and can be straightforwardly evaluated in the limit $\xi^2 \to 0$. So we have (in Minkowski metric) 
\begin{eqnarray}
&&{\cal M} (P\cdot \xi,0) = \langle P|\phi (0) \phi(\xi)|P\rangle \Big{|}_{\xi^2=0} = \int_{- \infty}^{+ \infty} d \omega \,f (\omega) e^{- i \omega P\cdot \xi} \label{bilocstru1} \\
&&f (\omega) = \frac {1}{2 \pi} \int_{- \infty}^{+ \infty}d (P\cdot \xi)  {\cal M}  (P\cdot \xi,0) \,e^{i \omega P\cdot \xi} \, .\label{bilocstru2}
\end{eqnarray}
Denoting by $W(q^2, q\cdot P)$ the DIS cross section, one gets~\cite{GROSS,Yndurain}
\begin{eqnarray}
&&(2 \pi)^4 W (q^2, q\cdot P) = \int_{- \infty}^{+ \infty} d \omega\, f (\omega) \int d^4 x \, e^{- i (q + \omega P) \cdot x} \Delta (x^2) = \nonumber \\
&&= (2 \pi)^4 \int_{- \infty}^{+ \infty} d\omega \,f(\omega) \delta [(q + \omega P)^2] \epsilon [ (q + \lambda P)_0]\, , \label{W}
\end{eqnarray}
where $\Delta(x^2)= \int \frac {d {\bf k}} {2 |{\bf k}|} e^{i k \cdot x} = \int d^4 k \, \delta (k^2) \epsilon (k^0)\, e^{ik\cdot x} $. Eq.~(\ref{W}) gives in the Bjorken limit  
\begin{eqnarray}
W (q^2, q\cdot P) \approx \frac {\omega f (\omega)}{-q^2} \, , \qquad \omega \equiv - \frac {q^2}{2 q \cdot P} \, .
\label{WOM}
\end{eqnarray}
The relation~(\ref{WOM}) expresses the DIS cross section in terms of the Fourier Transform (FT) of the matrix element of a bilocal operator (see eq.~(\ref{bilocstru2})).

The bilocal operator in eq.~(\ref{bilocstru1}) can be formally Taylor-expanded around $\xi_\mu=0$, yielding 
\begin{eqnarray}
\hspace{-.8cm}&&\langle P| \phi (0) \phi(\xi) |P \rangle \!= \!\sum_{n=0}^{\infty} \frac {1} {n!} \langle P|\phi(0) \frac{\partial}{\partial \xi^{\mu_1}}\frac{\partial}{\partial \xi^{\mu_2}}\ldots \frac{\partial}{\partial \xi^{\mu_n}}\phi (\xi)\Big{|}_{\xi=0}|P\rangle \xi^{\mu_1} \xi^{\mu_2} \dots \xi^{\mu_n} \!\equiv \!\nonumber\\
\hspace{-.8cm}&&\equiv \sum_{n=0}^{\infty} \langle P| O_{\mu_1\mu_2 \dots \mu_n} (0) |P \rangle \xi^{\mu_1} \xi^{\mu_2} \dots \xi^{\mu_n} \, .
\label{TAYLOR}
\end{eqnarray}
The matrix elements of the DIS operators, $O_{\mu_1 \mu_2\dots \mu_n} (0)$, are of the form
\begin{eqnarray}
\langle P| O_{\mu_1 \mu_2\dots \mu_n} (0) |P \rangle = A_n P_{\mu_1}P_{\mu_2} \dots P_{\mu_n} + traces \, , \label{locopbil}
\end{eqnarray}
where $traces$ denote terms containing some $g_{\mu_i \mu_j}$ tensor. The physical PDFs are related to the $A_n$ form factors (moments), while the $traces$ $B_n$ are spurious contributions which need to be subtracted out. In the Minkowski metric this subtraction is automatically performed by taking $\xi^2 = 0$ (as in eq.~(\ref{bilocstru1})). In Euclidean metric the situation is more complicated and is discussed in sect.~\ref{sec:RENMAT}.

\subsection{Comparing Light-Cone with Equal-Time structure function}
\label{sec;ETLC}

We now demonstrate that, neglecting renormalization issues, the Ji formula gives back the  LC structure function. The proof is carried out by showing that the formal spectral representations of the two quantities are equal. 

{\it Equal-Time structure function} - To compute the spectral representation it is convenient to work in a reference frame where the proton is at rest. The boost that brings the proton with four-momentum $P_\mu=(\sqrt{M^2+P^2},0,0,-P)$ at rest is given by the Lorentz transformation
\begin{eqnarray}
&&{x^0}' = \frac {x^0 - \beta z}{\sqrt{1-\beta^2}}\, , \qquad z' =  \frac {z - \beta x^0}{\sqrt{1-\beta^2}} \, , \quad {\mbox{with}} \quad \beta =- \frac {P} {\sqrt {M^2 +P^2}} \, .\label{LOR}
\end{eqnarray}
From the above relations, after inserting a complete set of states, we obtain for the ET bilocal 
\begin{eqnarray}
&& \langle P|\phi (0) \phi(\xi)|P \rangle\Big{|}_{\xi=(0,0,0,z)} = \langle M | \phi (0) \phi (\frac {P} {M} z, 0, 0, \frac {\sqrt {M^2 +P^2}}{M} z )|M \rangle = \nonumber \\
&&= \sum_n |\langle n|\phi(0)|M \rangle|^2 e^{i (E_n - M)  \frac {P} {M} z} e^{-i p_{n_z} \frac {\sqrt {M^2 +P^2}}{M} z} \, .\label{SPECT}
\end{eqnarray}
From the definition~(\ref {proposal}) we therefore get 
\begin{eqnarray}
\hspace{-.8cm}&&F (\omega) = \lim_{P \rightarrow +\infty} P\sum_n |\langle n|\phi(0)|M \rangle|^2 \delta \Big{(}\omega P + (E_n - M)  \frac {P} {M}- {p_n}_z \frac {\sqrt {M^2 +P^2}}{M}\Big{)} = \nonumber \\
\hspace{-.8cm}&&
= \sum_n |\langle n|\phi(0)|M \rangle|^2 \delta \Big{(}\omega - 1+ \frac {E_n - {{p_n}_z}} {M}\Big{)}\, . \label{ZFOR}
\end{eqnarray}

{\it Light-Cone  structure function} - Let $\xi$ be the light-like vector $(z,0,0,z)$. Using the definitions~(\ref{bilocstru1})-(\ref{bilocstru2}) and following a line of arguments similar to the ones employed above, we get
\begin{eqnarray}
\hspace{-.8cm}&&f (\omega) =  \frac {M} {2 \pi} \sum_n |\langle n | \phi(0) |M\rangle|^2 \int_{- \infty}^{+ \infty} d z \,e^{i z (-M+E_n- {p_n}_z+ \omega M)} 
=\nonumber \\&&
= \sum_n |\langle n | \phi(0) |M \rangle|^2 \delta \Big{(}\omega-1 + \frac {E_n- {p_n}_z}{M}\Big{)} \label{spectrum} \, ,
\end{eqnarray}
which indeed coincides with the last equality in eq.~(\ref{ZFOR}). Hence, barring renormalization effects, eq.~(\ref{proposal}) provides the ET version of eq.~(\ref{spectrum}).

\subsubsection{Disclaimer}
\label{sec:DISCL}

The correspondence between~(\ref{ETB}) and~(\ref{ELC}), or between~(\ref{ZFOR}) and~(\ref{spectrum}), is formal. We must observe, in fact, that if we work at ET we do not encounter in~(\ref{ETB}) the logarithmic singularities associated with anomalous dimensions but, as we shall see, the moments of the structure function, $F(\omega)$, are affected by UV power divergencies coming from mixing with lower dimensional ($trace$) operators. Vice versa, on the LC, UV power divergent mixings give raise to sub-leading space-time behaviours that will not affect the moments of $f(\omega)$, but the bilocal~(\ref{ELC}) will display singular anomalous dimension logarithms that must be appropriately dealt with in the small $\xi^2$ expansion.

\section{Renormalization and matching}
\label{sec:RENMAT}

In QCD violations of the Bjorken scaling in the DIS region are controlled by computable logarithmic corrections. The local operators in eq.~(\ref{locopbil}) require a renormalization which is not simply multiplicative, as they mix with UV power divergent coefficients, with lower dimensional ($trace$) operators. One needs to resolve this mixing to make the $A_n$ and $B_n$ form factors finite. In particular in order to be able to take the limit $P \to \infty$, necessary to eliminate the contamination from higher twists, one needs to make the $B_n$'s finite. However, the only renormalization considered in~\cite{Ji:2013dva,Lin:2017snn} and in most of the subsequent papers on the subject, was a multiplicative one. 

For concreteness the argument we develop here refers to the matching formula originally proposed by Ji, but we want to stress that the same kind of reasoning would apply to any logarithmic, multiplicative renormalization. The basic procedure, common to many of the approaches that have been following in a way or another the Ji idea~\cite{Ji:2013dva} is to start by  renormalizing the quantity ($\Lambda$ is the UV cutoff) 
\begin{eqnarray}
&&{\tilde F} (\omega, P;\Lambda) = \frac {P}{2 \pi} \int_{- \infty}^{+\infty} dz \, e^{i \omega z P} \langle P |\phi(0) \phi(\xi)|P\rangle \Big{|}^\Lambda_{\xi=(0,0,0,z)} \, ,
\label{fund}
\end{eqnarray}
to cure the bilocal logarithmic divergency. In particular the so-called ``matching condition'' consists in constructing the UV finite quantity $F(x,P; \mu)$ from the convolution formula
\begin{eqnarray}
 {\tilde F} (\omega, P; \Lambda) =  \int_\omega^{+\infty} \frac {d \omega'}{\omega'} Z (\frac {\omega}{\omega'}; \Lambda,\mu) F(\omega',P;\mu) \, ,
 \label{CONV}
\end{eqnarray}
where $Z ({\omega}/{\omega'}; \Lambda, \mu )$ is a logarithmically divergent renormalization factor, computable in PT. Taking moments of eq.~(\ref{CONV}) one immediately gets
\begin{eqnarray}
\hspace{-.8cm}&&\int_{-\infty}^{+\infty} d \omega \,\omega^n \, {\tilde F} (\omega, P; \Lambda) = \int_{-\infty}^{+\infty} d\omega' \,{\omega'}^{\,n} \, Z(\omega'; \Lambda,\mu) \int_{-\infty}^{+\infty} d\omega \, \omega^n \, F (\omega,P; \mu) \!\equiv \nonumber \\
\hspace{-.8cm}&&\equiv Z_n \left (\frac {\Lambda} {\mu} \right) \int_{-\infty}^{+\infty} d\omega \,\omega^n F (\omega,P; \mu)  \, , \label{mellconv}
\end{eqnarray}
implying that the moments of $\tilde F$ are separately multiplicatively renormalized. 

The key observation about eq.~(\ref{mellconv}) is is that it is expected to become a relation involving the moments of the physical PDF after sending $P\to \infty$. Taking this limit on the lattice is, however, not possible as we now show. In fact, eq.~(\ref{fund}) is a Fourier Transform, the inverse of which reads 
\begin{eqnarray}
\langle P |\phi(0) \phi({\xi})|P\rangle\Big{|}^\Lambda_{\xi=(0,0,0,z)} = \int_{-\infty}^{+\infty} d \omega \, e^{- i \omega  {z} P} {\tilde F}(\omega , P; \Lambda) \, . \label{genfun}
\end{eqnarray}
Computing the $n$-th derivative of~(\ref{genfun}) with respect to ${z}$ at ${z}=0$ gives 
\begin{eqnarray}
(-i)^n \int_{-\infty}^{+\infty} d \omega \, \omega^n {\tilde F}(\omega, P; \Lambda) = \frac {1}{(P)^n} \langle P |\phi(0) \frac {\partial^n \phi}{\partial z^n}(0)|P\rangle \Big{|}^\Lambda \, ,
\end{eqnarray}
which together with eq.~(\ref{mellconv}) implies
\begin{eqnarray}
\hspace{-1.cm}&&\int_{-\infty}^{+\infty} \!d\omega  \,\omega ^n  F (\omega , P; \mu)  = \frac {(-i)^n} {Z_n (\frac{\Lambda}{\mu})} \int_{-\infty}^{+\infty}\! d \omega  \,  \omega ^n {\tilde F}(\omega , P; \Lambda) = \frac {1}{(P)^n} \langle P | \frac {1} {Z_n (\frac{\Lambda}{\mu})} \phi(0) \frac {\partial^n \phi}{\partial z^n}(0)|P\rangle \Big{|}^\Lambda \!\, . \label{finedellastoria}
\end{eqnarray}
We see that the moments of the PDF derived from the multiplicatively renormalized eq.~(\ref{proposal}) are plagued by UV divergencies coming from the unsubtracted equal-point singularities of the operators $\phi(0) \frac {\partial^n \phi}{\partial z^n}(0)$. Thus lattice simulations of eq.~(\ref{proposal}) cannot give acccess to the physical PDF.

\subsection{An example: the second moment}
\label{sec:SECMOM}

The problem of eq.~(\ref{proposal}) with the mixing of $trace$ operators is best exhibited by explicitly computing PDF moments. If for illustration we consider the case of the second moment, we have  
\begin{eqnarray}
&&\int_{- \infty}^{+ \infty}d \omega \omega^2 F (\omega)  = \lim_{P \rightarrow +\infty} \frac {P}{2 \pi} \int_{-\infty}^{+\infty}  d \omega\, d z \,\omega^2 e^{iz \omega P} \langle M | \phi (0)  \phi (\frac {P}{M} z,0,0,\frac {\sqrt{M^2+P^2}}{M}z) |M\rangle =\nonumber \\
&&= - \lim_{P \rightarrow +\infty} \frac {1}{P^2} \int_{-\infty}^{+\infty} d z \frac {d^2 \delta (z)}{d z^2} \langle M |\phi (0) \phi (\frac {P}{M} z,0,0,\frac {\sqrt{M^2+P^2}}{M}z) |M \rangle = \nonumber \\
&&=- \lim_{P \rightarrow +\infty} \frac {1}{P^2} \frac {d^2}{d z^2} \langle M |\phi (0) \phi (\frac {P}{M} z,0,0,\frac {\sqrt{M^2+P^2}}{m}z)  |M\rangle \Big{|}_{z=0} \, .  \label{SECMOM} 
\end{eqnarray}
The connection between the second moment and the lowest rank-two local operator is therefore
\begin{eqnarray}
\hspace{-.2cm}&&\int_{- \infty}^{+ \infty} d \omega \,\omega^2 F (\omega)  = \label{moment} \\
\hspace{-.2cm}&&=- \lim_{P \rightarrow +\infty} \frac {1}{P^2} \Big{(}\frac {P^2} {M^2}  \langle M |O^{(2)}_{00}(0) |M \rangle +  \frac {M^2+P^2} {M^2}  \langle M |O^{(2)}_{33}(0) |M \rangle +2 \frac {P \sqrt{M^2+P^2}} {M^2}  \langle M |O^{(2)}_{03}(0) |M \rangle\Big{)}\nonumber \, ,
\end{eqnarray}
where formally $O^{(2)}_{\mu \nu} = \phi (0) \partial_\mu \partial_\nu \phi(0)$. Ignoring divergences everything works fine~\cite{Brandt:1972nw,Fritzsch:2003fg} and from $\langle P| O^{(2)}_{\mu \nu} |P \rangle = A^{(2)} P_\mu P_\nu + B^{(2)} g_{\mu \nu} $ ($g_{\mu\nu}$ is the Minkowski metric tensor) we would get 
\begin{eqnarray}
&&\int_{- \infty}^{+ \infty} d \omega\, \omega^2 F (\omega) =  -  \lim_{P \rightarrow +\infty} (A^{(2)} - \frac {B^{(2)}}{P^2}) = - A^{(2)} \label{momentcan}  \, .
\end{eqnarray}
But, referring to the situation one encounters in lattice simulations, we immediately see that the contribution from the mixing of $O^{(2)}_{\mu \nu} $ with the lower dimensional $trace$ operator $a^{-2}\phi(0)^2 g_{\mu\nu}$ to eq.~(\ref{momentcan}) gives raise to the power divergent term 
\begin{eqnarray}
\hspace{-1.cm}&&\int_{- \infty}^{+ \infty} d \omega\, \omega^2 F (\omega)  \Big{|}_{\rm trace\,operator}\propto -\frac {1}{a^2 P^2}  \Big{(}\frac {P^2} {M^2} -  \frac {M^2+P^2} {M^2}  \Big{)} =  \frac {1}{a^2P^2} \, . \label{DIVE}
\end{eqnarray}
As on the lattice the largest attainable momentum is O($a^{-1}$), the limit $P\to \infty$ can only be taken after sending $a\to 0$. But in this limit the r.h.s.\ of eq.~(\ref{DIVE}) blows up. Thus, unless we perform the appropriate non-perturbative subtraction of power divergent terms in each one of the matrix elements in eq.~(\ref{TAYLOR}), the $P\to\infty$ limit cannot be taken.

The existence of this difficulty is also signalled by a problem with the support of $F(\omega)$ which in the LC approach can be proved to be $|\omega|\leq 1$. On the contrary, trace terms are not related to the current-hadron scattering amplitude and therefore will give contributions for all values of $\omega$. Their subtraction is essential for the validity of the ET calculation of the PDF. 

\subsection{An observation}
\label{sec:OBS}
 
In order to clarify a delicate point that was not properly appreciated in the literature we would like to end this section by remarking that one thing is to say that finite DIS operator matrix elements can be extracted by fitting the short distance OPE of the properly renormalized lattice bilocal onto the continuum expansion. A completely different thing is to say that the moments of $F(\omega)$ in eq.~(\ref{proposal}) (or of its logarithmically renormalized version) are finite. As we have shown, they are not. So they are not the same quantities one extracts from the OPE of the lattice bilocal. This means that its FT does not yield the correct PDF.

\section{Reduced Ioffe-time distributions and perturbative subtraction}
\label{sec:RIT}

The difficulties outlined in the previous sections, preventing the direct calculation of the PDF on the lattice, also affect the strategy advocated in refs.~\cite{Orginos:2017kos,Zhang:2018ggy} where it is proposed to consider as a better UV behaved quantity the reduced Ioffe-time distribution~\cite{IOFFE} 
\begin{eqnarray}
\mathfrak{M}(P\, z,z^2)=\frac{{\cal M}(-P\,z,-z^2)}{{\cal M}(0,-z^2)}
\label{ORGI}
\end{eqnarray}
with ${\cal M}(-P\,z,-z^2)$ the bilocal~(\ref{ETB}). Since the ratio $\mathfrak{M}(P\,z,z^2)$ only differs from ${\cal M}(-P\,z,-z^2)$ by a ($z^2$-dependent) rescaling, the problem with power divergent moments is still present. 

From the small $z^2$ OPE of the lattice regularized ratio~(\ref{ORGI}) in terms of ``continuum'' Wilson coefficients (say in the $\overline{MS}$ scheme) one can in principle extract the correct (finite) PDF moments~\cite{Karpie:2018zaz}. However, in order to directly construct the whole PDF from lattice data one would need to take the FT of the quantity~(\ref{ORGI}). This FT will display power divergent moments irrespective of whether they are defined as derivatives of quasi-PDF's with respect to $z$ (at fixed $P$) or of pseudo-PDF's with respect to $\nu=P\, z$ (at fixed $z$). In the first case we are in the same situation as for the original Ji proposal (see our discussion in sect.~\ref{sec:RENMAT}). In the second, in order to take the derivatives with respect to $\nu=P\, z$ at vanishing $z^2$, one needs to send $P\to \infty$, which is impossible in the lattice regularization. 

As a way to circumvent these problems, the Authors of ref.~\cite{Orginos:2017kos} have proposed to subtract out the unwanted terms in PT. We illustrate their idea and the difficulties that go along with it with the help of the illuminating approach and notations of ref.~\cite{Radyushkin:2018nbf}.  

Using, say, dimensional regularization and the $\overline {MS}$ subtraction scheme, the regularized quasi-PDF, $Q(\omega,P)$, can be related in PT to the LC continuum PDF, $f(\omega;\mu^2)$, by the formula~\cite{Zhang:2018ggy,Radyushkin:2018nbf}
\begin{eqnarray}
\hspace{-.6cm}&&Q(\omega,P)=f(\omega;\mu^2) -\frac{\alpha_s}{2\pi}C_F\int_0^1\frac{du}{u}f(\frac{\omega}{u}; \mu^2)\Big{[}B(u)\ln(\frac{\mu^2}{P^2})+C(u)\Big{]}+\nonumber\\
\hspace{-.6cm}&& + \frac{\alpha_s}{2\pi}C_F\int_{-1}^1 dx\, f(x;\mu^2) L(\omega,x) +{\mbox{O}}(P^{-2}) +{\mbox{O}}(\alpha_s^{2})
\, ,\label{RAD1}
\end{eqnarray}
where $C(u)$ is a computable function, the explicit expression of which is not needed here and
\begin{eqnarray}
\hspace{-.6cm}&&L(\omega,x) =\frac{P}{2\pi}\int_0^1 du\, B(u) \int_{-\infty}^{+\infty} dz \,e^{-i(\omega-ux)zP} \ln (z^2P^2)\, .\label{RAD2}
\end{eqnarray}
The last term in eq.~(\ref{RAD1}) produces (unwanted) contributions in the $|\omega|>1$ region, yielding UV power divergent moments. One can thus think of subtracting out by hand these terms and write
\begin{eqnarray}
\hspace{-.8cm}&&f(\omega;\mu^2)= \Big{[}{Q(\omega,P)}- \frac{\alpha_s}{2\pi}C_F\int_{-1}^1 dx\, f(x;\mu^2) L(\omega,x) \Big{]}+\nonumber\\
\hspace{-.8cm}&& +\frac{\alpha_s}{2\pi}C_F\int_0^1\frac{du}{u}f(\frac{\omega}{u};\mu^2)\Big{[}B(u)\ln(\frac{\mu^2}{P^2})+C(u)\Big{]}
+{\mbox{O}}(P^{-2}) +{\mbox{O}}(\alpha_s^{2})\, .\label{RAD3}
\end{eqnarray}
The difficulties posed by this procedure, which is widely used in actual simulations, are as follows. First of all, we observe that the subtraction needs to be done before removing the cutoff. So all the formulae above should be looked at with this in mind. For instance, in lattice QCD simulations eq.~(\ref{RAD1}) and the following ones should be rewritten by using the lattice regularization.

Secondly, although it is true that the term in square parenthesis has a smooth $P\to \infty$ limit, the ${\mbox{O}}(\alpha_s^{2})$ corrections don't and at small lattice spacings they will matter. Indeed, UV power divergencies in moments are not eliminated but only pushed to higher orders in PT. 

Finally the very same PDF, $f(y;\mu^2)$, one is looking for appears in the r.h.s.\ of~(\ref{RAD3}). In practice to leading order in $\alpha_s$ one replaces it with the lattice quantity $Q(y,P)$. But the latter does not have the correct support properties. One thus needs to enforce them by hand. As a result non-localities are introduced. Then the question arises whether the moments of the PDF built in this way are the matrix elements of the renormalized local DIS operators~(\ref{locopbil}) one finds in the Bjorken limit. 

\section{Conclusions}
\label{sec:CONC} 

In this talk we have briefly rediscussed the viability of the proposal of directly extracting PDF's from QCD simulations. Unfortunately there is still a missing ingredient in this program, related to the problem of subtracting UV power divergent trace terms. Although finite, individual PDF moments can be extracted from lattice data, at the moment neither the initial Ji proposal~\cite{Ji:2013dva} of exploiting the formula~(\ref{proposal}), nor the direct use of the current-current $T$-product~\cite{Ma:2017pxb} or of the reduced Ioffe-time distributions~\cite{Orginos:2017kos,Zhang:2018ggy} allow to access the full PDF from lattice simulations. 

\appendix 
\section{Appendix - Trace operators in a toy-model}
\label{sec:APPB} 

To provide an intuition of the harm that UV power divergent mixings can cause in the construction of the PDF, we discuss a simple mathematical example mimicking what happens if divergent trace operators are not properly subtracted out in the process of renormalizing the leading twist local operators. Let us take as an explicit  toy-model for the matrix element of the ET bilocal the regularized expression 
\begin{equation}
\langle P| \phi(0) \phi(\xi)|P\rangle\Big{|}^\Lambda_{\xi=(0,0,0,z)} \to \,G(P \,z, z;\Lambda)=\int dk\, e^{-\frac{k^2}{\Lambda^2}}e^{ikz}g(P\, z,k) \, .
\label{GREP}
\end{equation}
The exponential factor $e^{ikz}$ has been introduced to describe the effects of trace operators. In fact, if Taylor-expanded, it gives rise to power divergent terms of the kind $(\Lambda z)^n$. In this model the matrix element of the properly subtracted leading twist operators is then obtained by just crossing out the $e^{ikz}$ factor from the the previous equation. If we do so, we get for the corresponding PDF 
\begin{eqnarray}
\hspace{-1.2cm}&& f(\omega;\Lambda)\! =\! P\! \int \! dz\,e^{i\omega P\, z} \! \!\int \! dk\, e^{-\frac{k^2}{\Lambda^2}}g(P\, z,k)\!=\!\!\int \! dk\, e^{-\frac{k^2}{\Lambda^2}} \,\tilde g(\omega,k) \, ,\,\, {\mbox{with}}\,\, \,\tilde g(\omega,k)\!=\! \!\int \!dy \,e^{i\omega y} g(y,k)\, .
\label{FOMEGAKT}
\end{eqnarray}
Eq.~(\ref{GREP}) leads instead to the PDF
\begin{eqnarray}
\widehat f(\omega;\Lambda) =P\int_{-\infty}^{\infty}dz\,e^{i\omega P\, z}  \int dk\, e^{-\frac{k^2}{\Lambda^2}}e^{ikz}g(P\, z,k) =\int dk\, e^{-\frac{k^2}{\Lambda^2}} \tilde g\Big{(}\omega+\frac{k}{P},k)\Big{)}\, .
\label{FOMEGAK}
\end{eqnarray}
We thus see  that mixings with trace operators do not show up as (power) divergencies in~(\ref{FOMEGAK}). Rather at finite $P$, they deform the expression of the latter compared to eq.~(\ref{FOMEGAKT}). Unfortunately in a regularized theory one cannot send $P$ to infinity as $P$ can never be made larger than $\Lambda$.

To complete the analysis we nned prove that, if the operators~(\ref{TAYLOR}) are made finite with the proper subtractions of the power divergent trace operators, the remaining, finite trace operator contributions to the structure function do indeed vanish in the limit of large $P$. 

In our toy-model the situation in which power divergent trace operator mixings are subtracted out from the bare operators~(\ref{TAYLOR}) can be mimicked by stipulating that the function $g(P\, z,k)$ has a well convergent behaviour for large $k$, say, with an exponential cutoff scaled by some physical, finite mass parameter, $\Lambda_s$. Thus, assuming for $g(P\, z,k)$ the behaviour~\footnote{For the sake of this argument one might equivalently well assume a behaviour of the kind $g(P\, z,k)\sim e^{-\frac{k^2}{\Lambda_s^2}} \hat h (P\, z,k)$.}
\begin{equation}
g(P\, z,k)\sim e^{-\frac{|k|}{\Lambda_s}} h(P\, z,k)
\label{GTO}
\end{equation}
with $h(P\, z,k)$ a smooth, bounded function of $k$, we can safely send the UV cutoff, $\Lambda$, to infinity in eq.~(\ref{GREP}), as the $k$ integral is convergent. In this situation one gets 
\begin{eqnarray}
\hspace{-.8cm}&&P\int_{-\infty}^{\infty}dz\,e^{i\omega P\, z}  \int dk\, e^{ikz} e^{-\frac{|k|}{\Lambda_s}}h(P\, z,k) \stackrel{P\gg\Lambda_s}\longrightarrow \int dk\, e^{-\frac{|k|}{\Lambda_s}} \tilde h\Big{(}\omega+\frac{k}{P},k\Big{)}=\int dk\,\tilde g(\omega,k)\, .
\label{FOMEGATR}
\end{eqnarray}
We see that the last expression coincides with eq.~(\ref{FOMEGAKT}), after removing there the UV cutoff. This last step can be safely performed as UV divergent trace operator mixings have been taken care of.

\end{document}